\documentclass[Journal]{IEEEtran}

\IEEEoverridecommandlockouts

\usepackage{bm}
\usepackage{color}
\usepackage{graphicx}
\usepackage{amsmath}
\usepackage{amssymb}
\usepackage{algorithm}
\usepackage{algorithmic}
\usepackage{ulem}
\usepackage{lineno}
\usepackage{lettrine}
\usepackage{subfigure}
\usepackage{epstopdf}
\usepackage{stfloats}
\usepackage{cite}
\usepackage{color}
\usepackage{graphicx}
\usepackage{amsmath}
\usepackage{amssymb}
\usepackage{algorithm}
\usepackage{algorithmic}
\usepackage{amsmath}
\usepackage{multirow}
\usepackage{booktabs}
\usepackage{array}
\usepackage{amsthm}
\usepackage{stfloats}
\usepackage{caption}
\usepackage{bm}
\usepackage{booktabs}
\usepackage{setspace}
\usepackage{makecell}
\usepackage{diagbox}

\newcommand{\be}{\begin{equation}}
\newcommand{\ee}{\end{equation}}
\newcommand{\bea}{\begin{eqnarray}}
\newcommand{\eea}{\end{eqnarray}}
\newcommand{\ba}{\begin{array}}
\newcommand{\ea}{\end{array}}

\title{Deep Learning based Efficient Symbol-Level Precoding Design for MU-MISO Systems}
\author{\IEEEauthorblockN{Zhu Bo, Rang Liu, \textit{Graduate Student Member}, \textit{IEEE}, Ming Li, \textit{Senior Member}, \textit{IEEE}, \\ and Qian Liu, \textit{Member}, \textit{IEEE}}\thanks{Z. Bo, R. Liu, and M. Li are with the School of Information and Communication Engineering, Dalian University of Technology, Dalian 116024, China (e-mail: zhubo@mail.dlut.edu.cn; liurang@mail.dlut.edu.cn; mli@dlut.edu.cn).

Q. Liu is with the School of Computer Science and Technology, Dalian University of Technology, Dalian 116024, China (e-mail:qianliu@dlut.edu.cn).}}

\begin{document}
\pagestyle{empty}
\maketitle
\thispagestyle{empty}
\begin{abstract}
The recently emerged symbol-level precoding (SLP) technique has been regarded as a promising solution in multi-user wireless communication systems, since it can convert harmful multi-user interference (MUI) into beneficial signals for enhancing system performance. However, the tremendous computational complexity of conventional symbol-level precoding designs severely hinders the practical implementations.
In order to tackle this difficulty, we propose a novel deep learning (DL) based approach to efficiently design the symbol-level precoders. Particularly, in this correspondence, we consider a multi-user multi-input single-output (MU-MISO) downlink system. An efficient precoding neural network (EPNN) is introduced to optimize the symbol-level precoders for maximizing the minimum quality-of-service (QoS) of all users under the power constraint. Simulation results demonstrate that the proposed EPNN based SLP design can dramatically reduce the computing time at the price of slight performance loss compared with the conventional convex optimization based SLP design.
\end{abstract}


\section{Introduction}
Multi-input multi-output (MIMO) technology has been widely studied in both academia and industry during the last decade, owing to its capabilities in increasing the capacity of wireless communication systems and providing sufficient beamforming gains to combat severe path loss \cite{MIMO}.
In order to enjoy these benefits, various precoding algorithms have been proposed under different requirements and goals.
Particularly, in multi-user MIMO (MU-MIMO) systems, the signals of interest are enhanced and the multi-user interference (MUI) is suppressed as much as possible via the traditional block-level precoding technique, which optimizes the precoders based on the channel state information (CSI) and the second-order statistics of the signals.

Unlike the interference suppression strategy of block-level precoding designs, the recently emerged symbol-level precoding (SLP) technique utilizes the transmitted symbolic information to convert the harmful MUI into beneficial signals.
In particular, the SLP technique utilizes MUI to push the received symbols (e.g. $M$ phase shift keying) further away from decision boundaries.
Thus, the system performance, e.g., the symbol error rate (SER) performance, can be significantly improved \cite{SLP1}.
The authors in \cite{SLP2} investigated constructive interference approaches on power minimization, max-min fairness and maximizing sum-rate problems.
In \cite{SLP3} the SLP technique was utilized in IRS-enhanced MISO systems and the authors proposed a joint SLP and reflecting algorithm which has remarkably performance on power-savings and SER-reductions.
A antenna selection based SLP approach was proposed for low power transceiver design in \cite{SLP4}.
However, the traditional SLP designs have inevitably huge computational complexity due to the high dimension of optimizing variable, which contains all possible transmitted precoding signals.
In order to employ SLP in practical time-varying wireless communications, it is urgent to explore breakthrough precoding design algorithms to radically reduce the computational complexity.

%%% DL
Fortunately, machine learning has been considered as a promising candidate solution to satisfy the diverse requirements of the next-generation communication systems \cite{DLA}.
Specifically, deep learning (DL) based approaches, which have many success stories in various disciplines, bring more possibilities to handle the optimization problems in wireless communication systems with much lower complexity.
The DL based approaches have been applied for addressing different issues in wireless communication systems. For example, in \cite{DLA1}, a DL based framework was proposed for fast power-allocation design.
The authors in \cite{DLA2,DLA3} exploit learning-based schemes for more accurate channel estimation.
%%%%
%%% DL BL
The DL-based block-level precoding designs have been widely investigated in \cite{DLBLP1,DLBLP2,DLBLP4,DLBLP5,DLBLP6,My}, which utilize various methods to convert the original optimization problems into DL training problems.
For example, the authors in \cite{DLBLP1} proposed a supervised convolutional neural network (CNN) to learn the mapping function from CSI to beamformers obtained by a conventional algorithm.
In \cite{DLBLP2}, the authors formulated the analog beamforming and combining design as a multi-label classification problem for DL network.
The popular supervised learning based block-level precoding approaches were also investigated in \cite{DLBLP4}.
In \cite{DLBLP5}, an unsupervised DL network was utilized to directly seek the optimal solutions of weighted-sum-rate and obtained better performance than the supervised scheme.
Unsupervised-like DL networks were also utilized in \cite{DLBLP6,My} to directly design precoders without the usage of labels.
%%% DL SL
The application of DL on symbol-level precoding was also investigated in \cite{DLSLP1}, where the authors proposed an auto-encoder based DL network for robust symbol-level precoding and detection design.
While the approach in \cite{DLSLP1} validates the feasibility of applying DL based methods for low-complexity symbol-level precoding designs, the complicated decision rules at the receivers using the proposed end-to-end network are quite difficult to be implemented in practice.

Motivated by these findings, in this correspondence we propose an efficient DL based SLP design with a straightforward decision rule, which thus brings significantly reduced complexity at both transmitter and receiver sides.
In particular, we aim to design the symbol-level precoders to maximize the minimum quality-of-service (QoS) of all users under given transmit power budget.
An unsupervised-like loss function is developed to train the proposed efficient precoding neural network (EPNN), which can rapidly provide the required symbol-level precoders.
Simulation results illustrate that the proposed EPNN dramatically reduces the computing time without causing much performance loss compared with the conventional SLP design algorithm.

\section{System Model and Problem Formulation }
\label{sc:system model}
We consider a downlink multi-user multi-input single-output (MU-MISO) communication system, where a base station (BS) equipped with $N_{\mathrm{t}}$ antennas serves $K$ single-antenna users.
We assume that the transmitted signals are independently selected from $M$ phase shift keying ($M$-PSK) modulation, e.g., $M$ = 2, 4, 8, etc.
Thus, the transmitted symbol vector $\mathbf{s}_l \in \mathbb{C}^K$ for all $K$ users has $M^K$ different combinations, $l = 1, 2, \ldots, M^K$.
For transmitting a certain symbol vector $\mathbf{s}_l$, the symbol-level precoder $\mathbf{x}_l \in \mathbb{C}^{N_\mathrm{t}}$ is designed correspondingly and then transmitted via $N_{\mathrm{t}}$ antennas. Thus, the received signal at the $k$-th user is written as:
\begin{equation}
r_{k,l}=\mathbf{h}_k^H \mathbf{x}_{l}+n_k, \forall k, \forall l,
\end{equation}
where $\mathbf{h}_k \in \mathbb{C}^{N_\mathrm{t}} $ is the channel vector between the BS and the $k$-th user, and $n_k \sim \mathcal{CN}(0, \sigma^2_k)$ is the additive white Gaussian noise (AWGN) at the $k$-th user.
%\begin{figure}
%\centering
%\includegraphics[width=7.2cm]{Detection3.eps}%H_a.eps
%\caption{Symbol detection operation.}
%\label{2}
%\vspace{-0.6 cm}
%\end{figure}
\begin{figure}[!t]
\centering
\subfigure[An example of detection region.]{\includegraphics[width= 1.9 in]{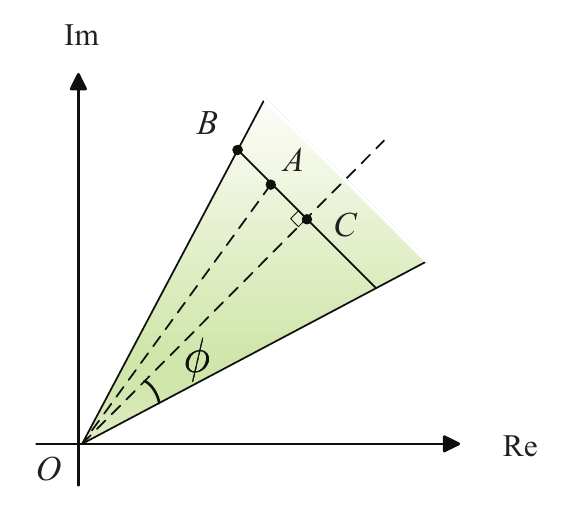}} \vspace{0.3 cm}
\subfigure[After rotating the diagram in Fig. 1(a) clockwise.]
{\includegraphics[width= 2.3 in]{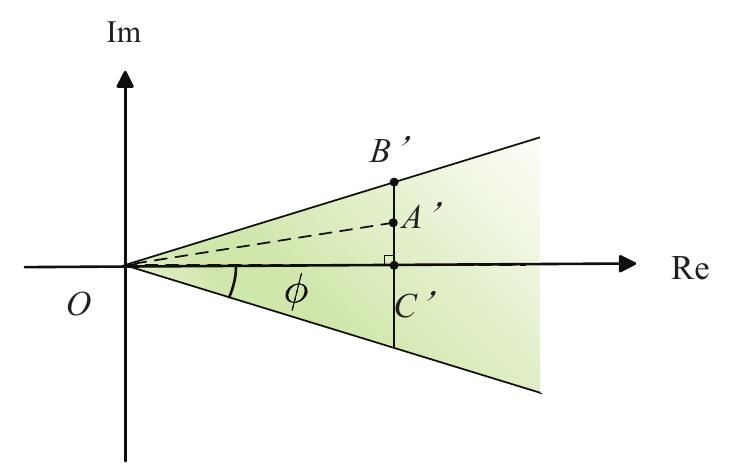}}

\caption{Symbol detection operation.}

\end{figure}
To formulate the QoS metric for SLP, we show in Fig. 1 the constellation at the $k$-th user when $\mathbf{x}_l$ is transmitted, where point $A$ is the received noise-free signal, i.e., $\overrightarrow{OA} = \widetilde{r}_{k,l} = \mathbf{h}^H_k\mathbf{x}_l$.
The solid black lines are the decision boundaries of the desired symbol $\mathbf{s}_l(k)$ and the black dotted line is the angle bisector of the green region.
The angle of the decision region is $2\times\phi$, $\phi=\frac{\pi}{M}$.
Point $C$ is the projection of $A$ on the angle bisector.
Point $B$ is the intersection of $\overrightarrow{CA}$ and the nearest decision boundary.
%We see that if ${r}_{k,l}$ is in the green region, the symbol can be correctly demodulated.
The decision areas are angle-dependent sectors, e.g., the green region in Fig. 1. We see that if $r_{k,l}$ is in the green region, the received symbol can be correctly demodulated as the desired symbol $\mathbf{s}_l(k)$.
In order to improve the robustness to the noise, the received noise-free signal, i.e., $\overrightarrow{OA}$, should be as farther away from its decision boundaries as possible.
Therefore, the distance between the received noise-free signal and its nearest decision boundary, which is expressed as $(|\overrightarrow{CB}|-|\overrightarrow{CA}|)\cos\phi$, is regarded as the QoS metric to evaluate the SER performance.

To explicitly express the QoS metric, we rotate the diagram in Fig. 1(a) clockwise by $\angle \mathbf{s}_l(k)$ degrees as shown in Fig. 1(b).
The QoS metric is thus expressed as\footnote{More detailed descriptions about symbol-level precoding can be found in \cite{SLP1,SLP2,SLP3,SLP4}.}:

\begin{equation}
d_{k,l}=(\Re\{{\mathbf{h}_k^H \mathbf{x}_{l}e^{-j\angle \mathbf{s}_l(k)}}\}\tan\phi-|\Im\{\mathbf{h}_k^H \mathbf{x}_{l}e^{-j\angle \mathbf{s}_l(k)}\}|)\cos \phi.
\end{equation}
Larger $d_{k,l}$ indicates that the received noise-free signal is farther away from its decision boundaries, which produces smaller SER.
Therefore, the max-min fairness problem is formulated as:
\begin{equation}
\begin{aligned}
&\underset{\mathbf{X}}{\max}~~\underset{k,l}{\min}~~d_{k,l,}\\
&\mathrm{s.t.}~~ \hspace{-0cm}
\|\mathbf{X}\|^2_F\leq PM^K, \\
\end{aligned}
\label{PRO}
\end{equation}
where $\mathbf{X}\triangleq[\mathbf{x}_1,\mathbf{x}_2,...,\mathbf{x}_{M^K}]$ is the precoding matrix for all possible transmitted symbol vectors, and $P$ is average transmit power budget.

Problem (\ref{PRO}) is convex and can be solved by convex optimization toolbox, e.g., CVX. However, the traditional convex based optimization algorithms lead to huge computational complexity due to the following reasons: 1) The popular iterative algorithms usually require considerable iterations;
2) The dimension of variable $\mathbf{X} \in \mathbb{C}^{N_\mathrm{t}\times M^K}$ exponentially increases with the number of users.
Although the symmetrical feature of constellation can be utilized to reduce the dimension of the optimizing variable to $N_\text{t} \times N_\text{par}$, $N_\text{par} = M^{K-1}$ \cite{Rotate}, the computational complexity is still unaffordable with large numbers of users and antennas.
\begin{figure*}
\centering
\includegraphics[height=5.0cm]{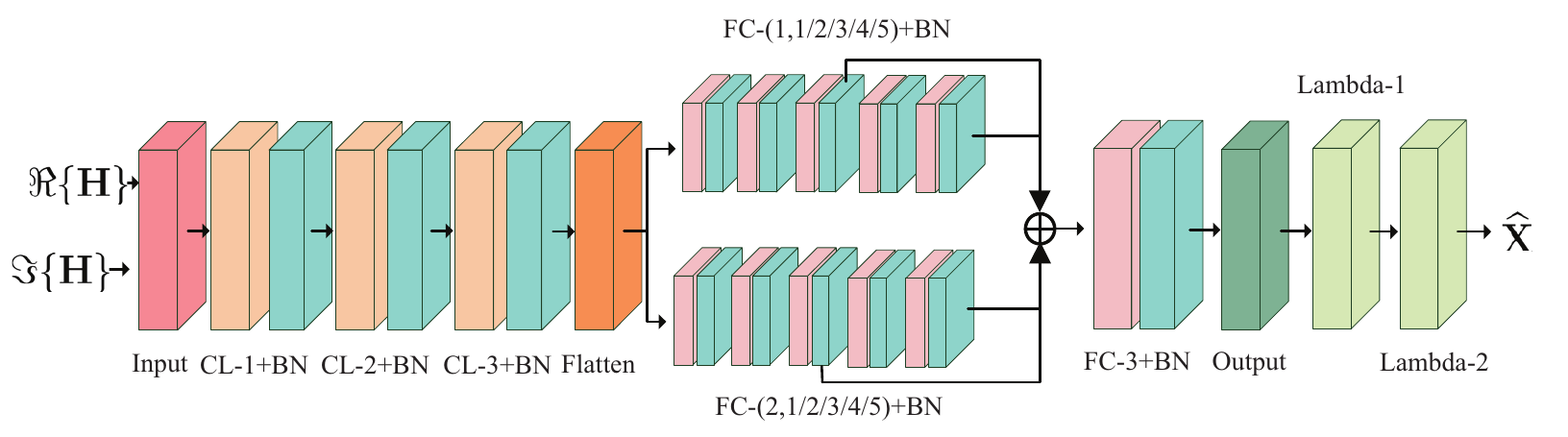}%H_a.eps
\caption{The structure of efficient precoding neural network (EPNN).}
\label{2}
\end{figure*}

It is worth noting that symbol-level precoder $\mathbf{x}_l$ can essentially be written as a function of the CSI and the symbol vector $\mathbf{s}_l$:
\begin{equation}
\mathbf{x}_l=\mathcal{F}(\mathbf{H},\mathbf{s}_l),
\end{equation}
where $\mathcal{F}(\cdot)$ represents a non-linear function, $\mathbf{H}\triangleq[\mathbf{h}_1,\mathbf{h}_2,...,\mathbf{h}_K] \in
\mathcal{}\mathbb{C}^{N_\mathrm{t} \times K}$.
Fortunately, deep neural networks are proficient in solving for this  non-linear mapping problem, thanks to the non-linearity introduced by activation functions.
Therefore, in order to dramatically reduce the complexity while maintain a satisfactory performance, we propose an efficient deep learning based approach to solve the SLP design problem (\ref{PRO}).
The trained deep learning network can construct a directly mapping function from $\mathbf{H}$ and $\mathbf{s}_l$ to the desired $\mathbf{x}_l$.
In addition, we should emphasize that the information symbols can be easily demodulated at the receiver side via a simple hard-decision detector, since the optimization problem (3) assumes angle-dependent sector decision areas. Compared with the irregular decision areas utilized in \cite{DLSLP1}, our proposed method is very appealing in practical hardware deployment.

\section{DL-based Symbol-Level Precoder Design}
In this section, we propose an efficient precoding neural network, which is referred to as EPNN, to design the symbol-level precoder in an unsupervised-like approach.
The structure and training strategy of EPNN are described in the following subsections, respectively.

\subsection{Structure of EPNN}
The structure of EPNN  is shown in Fig. 2, where the input is CSI matrix $\mathbf{H}$ and the output is the precoding matrix $\widehat{\mathbf{X}}\triangleq[\hat{\mathbf{x}}_1,\hat{\mathbf{x}}_2,...,
\hat{\mathbf{x}}_{N_\mathrm{par}}]$.
Since DL networks cannot handle complex data, we extract the real and imaginary parts of $\mathbf{H}$, i.e., $\Re(\mathbf{H})$ and $\Im(\mathbf{H})$, as the input.
The main structure of the network consists of three convolutional (CL) layers and 11 fully-connected (FC) layers.
At the end of this network, there are two non-trainable Lambda layers for reshaping and normalizing operations.
The CL layers extract the features of input data and the FC layers utilize these features for regression.
The detailed description is as follows.

The input data is first processed by the CL layers and vectorized by the Flatten layer, where we adopt ``SAME'' padding. The stride of CL-1 layer is set as ($K$,1) in order to focus on the relationship among different channels, and the strides of the other CL layers are set as (1,1) to capture more detailed features since the dimension of input data is small.
%and choose strides as (1,1) except CL-1, since the dimension of input data is small.
%The stride of CL-1 layer is set as ($K$,1) to capture the relationship between different user channels.
Then, a parallel structure of FC layers is utilized for obtaining more information, where we employ a residual connection between FC-(1/2,3) and the end of the parallel structure.
The residual connection provides more flexibility to the network and mitigates the problem of gradient vanishing in deeper parts of the network.
We adopt the rectified linear unit (ReLU) as activation function followed by each CL layer and FC layer.
The ReLU function, represented by $\sigma_{\mathrm{ReLU}}(x)=\max(0,x)$, is a non-linear function that enables the network to handle non-linear problems.
Besides, the batch normalization (BN) layer is added after ReLU function to prevent overfitting and accelerate training process.
The output data of output layer has a dimension of $2N_\mathrm{t}N_\mathrm{par}\times 1$ and contains both real and imaginary parts of $\widehat{\mathbf{X}}$.

At the end of this network, two Lambda layers are applied for reshaping, converting and scaling operations.
After the output layer, Lambda-1 layer is applied to reshape and convert the real-valued output parts of $\widehat{\mathbf{X}}$ to the complex valued $\widehat{\mathbf{X}}_\mathrm{temp} \in \mathbb{C}^{N_\mathrm{t}\times N_\mathrm{par}}$.
%Specifically, Lambda-1 layer reshapes the data of output layer to $\widehat{\mathbf{X}}_\mathrm{temp} \in \mathbb{C}^{N_\mathrm{t}\times N_\mathrm{par}}$.
Then, we can obtain precoders under different power constrains by scaling operation.
Lambda-2 layer scales $\widehat{\mathbf{X}}_\mathrm{temp}$ as:
\begin{equation}
\mathbf{\widehat{X}}=\frac{\mathbf{\widehat{X}}_{\mathrm{temp}}}
{\|\mathbf{\widehat{X}}_{\mathrm{temp}}\|_{\mathrm{F}}}
\sqrt{\min(\|\mathbf{\widehat{X}}_{\mathrm{temp}}\|_{\mathrm{F}},P N_\mathrm{par})}.
\end{equation}
The parameters of the proposed EPNN are summarized in Table I.
\begin{table}
\normalsize
\caption{Parameters of proposed EPNN}
\begin{center}
\begin{tabular}{|c|c|}
\hline
Type of layer & Parameters\\
\hline CL-1 & $256\times(K,1)$\\
\hline CL-2,3 & $256\times(1,1)$\\
\hline FC-(1/2,1/2) & 2048 \\
\hline FC-(1/2,3) & 8192\\
\hline FC-(1/2,4) & 2048 \\
\hline FC-(1/2,5) & 8192 \\
\hline FC-3 & 2048 \\
\hline Output & $2{N_{\mathrm{t}}}N_{\mathrm{par}}$ \\
\hline
\end{tabular}
\end{center}
\end{table}

\subsection{Training Strategy of EPNN}
In this subsection, we propose an unsupervised-like training strategy with the corresponding loss function for the off-line training stage.
It is noted that the supervised learning approaches require the solutions obtained by the non-DL algorithms as labels to train the network by minimizing the mean square error (MSE) between the obtained solution and the labels.
However, as stated before, using the convex optimization algorithms to solve problem (\ref{PRO}) is very time-consuming.
Furthermore, the MSE metric probably causes performance loss since it is irrelevant to the original optimization objective.
Thus, we adopt the unsupervised-like training approach for the SLP design to avoid the challenge task of generating a large number of labeled data.
The loss function of the proposed unsupervised-like training approach is formulated as:
\begin{align}
\mathcal{L}_{\mathrm{loss}}&= -\nu
+\frac{1}{\lambda}\frac{1}{KN_\mathrm{par}}\sum_{k=1}^{K}\sum_{l=1}^{N_\mathrm{par}}(\nu-\hat{d}_{k,l})^2,
\label{loss}
\end{align}
where $\lambda$ is a regularization factor, $\nu$ and $\hat{d}_{k,l}$ are defined as:
\begin{align}
\nu&\triangleq\frac{1}{KN_\mathrm{par}}\sum_{k=1}^{K}\sum_{l=1}^{N_\mathrm{par}}\hat{d}_{k,l},\\
\hat{d}_{k,l}&\triangleq( \Re\{
\mathbf{h}^H_{k} \mathbf{\hat{x}}_{l} e^{-j\angle \mathbf{s}_l(k)}
\}\tan\phi
-|\Im\{\mathbf{h}^H_{k} \mathbf{\hat{x}}_{l} e^{-j\angle \mathbf{s}_l(k)}  | ) \cos \phi.
\end{align}
The loss function (\ref{loss}) is a variation of the objective in original optimization problem (\ref{PRO}), where the minimum $d_{k,l}$ is maximized to guarantee fairness.
In the network training process, directly employing the objective of (\ref{PRO}) will cause excessive focuses on the worst precoder while neglecting the features of the whole precoding matrix $\widehat{\mathbf{X}}$, which will interrupt the training process at a poor local optimum.
Therefore, in order to balance the overall performance and fairness, we propose to utilize (\ref{loss}) as the loss function, which contains the average value and the regularized variance of $d_{k,l}, \forall k,l$.
It is worth noting that our goal is utilizing instantaneous CSI to generate symbol-level precoders for all possible combinations of transmitted symbols.
The information of different transmitted symbol vectors is unnecessary for input, since they are pre-defined non-trainable hyper-parameter of EPNN and have been taken into consideration in the loss function (\ref{loss}).

\vspace{0.2cm}
\section{Simulation Results}
In this section, we present the simulation results to illustrate the effectiveness of our proposed EPNN approach in SLP design.
We assume that the system adopts quadrature phase shift keying (QPSK) modulation and the CSI is perfectly known to the BS.
All the channels follow a classical Rayleigh fading model, i.e., all entries of $\mathbf{H}$ follow the distribution $\mathcal{CN}(0, 1)$.
TensorFlow is utilized for the employment of the EPNN framework.
All simulations are deployed on a computer with NVIDIA GeForce GTX 1660 GPU and Intel(R) Core(TM) i7-8700 CPU.
\begin{figure}[!t]
\centering
\subfigure[$N_{\mathrm{t}}=4$, $K=3$]{\includegraphics[width= 3.3 in]{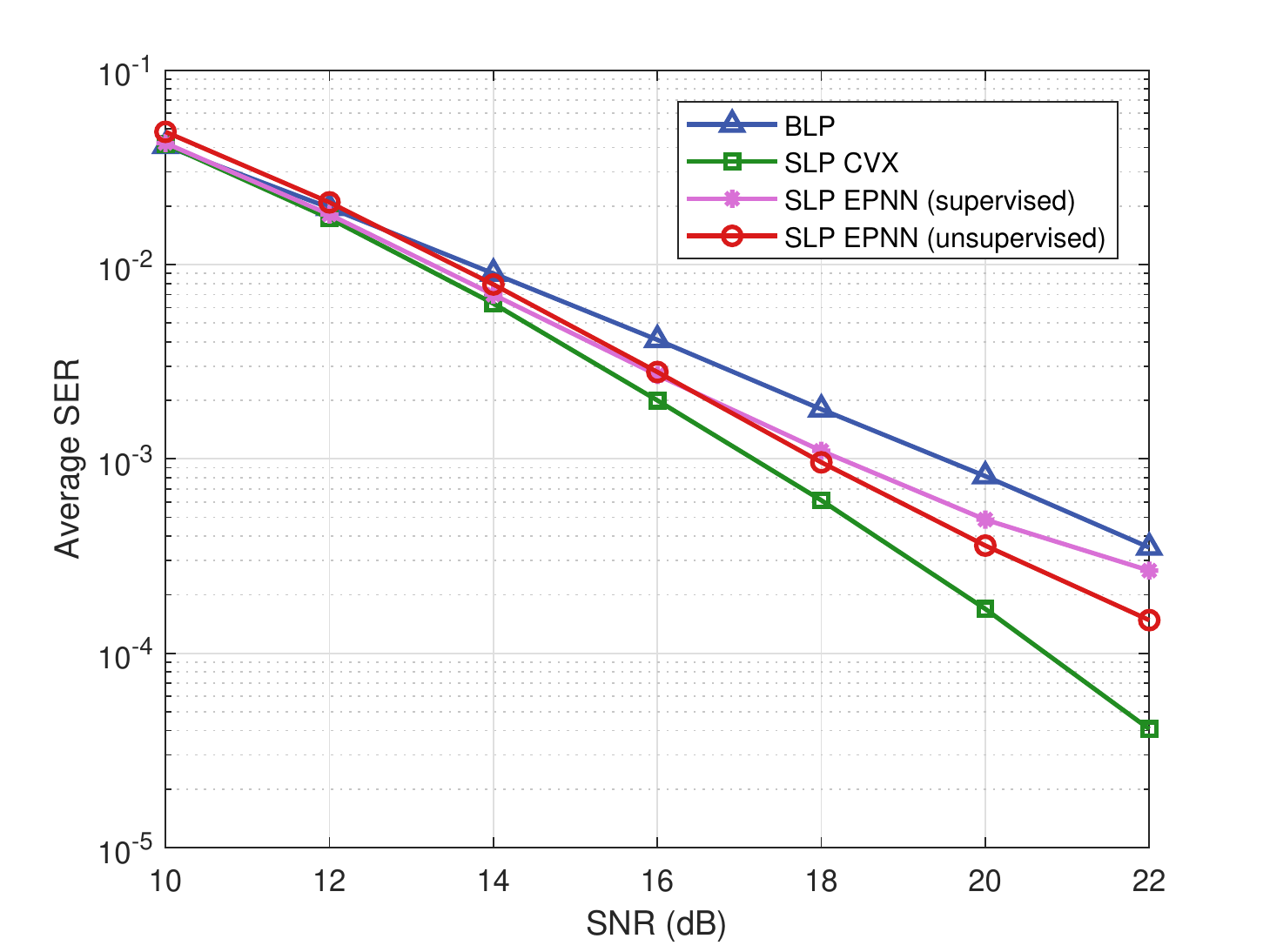}} \vspace{-0.4 cm}
\subfigure[$N_{\mathrm{t}}=5$, $K=4$]{\includegraphics[width= 3.3 in]{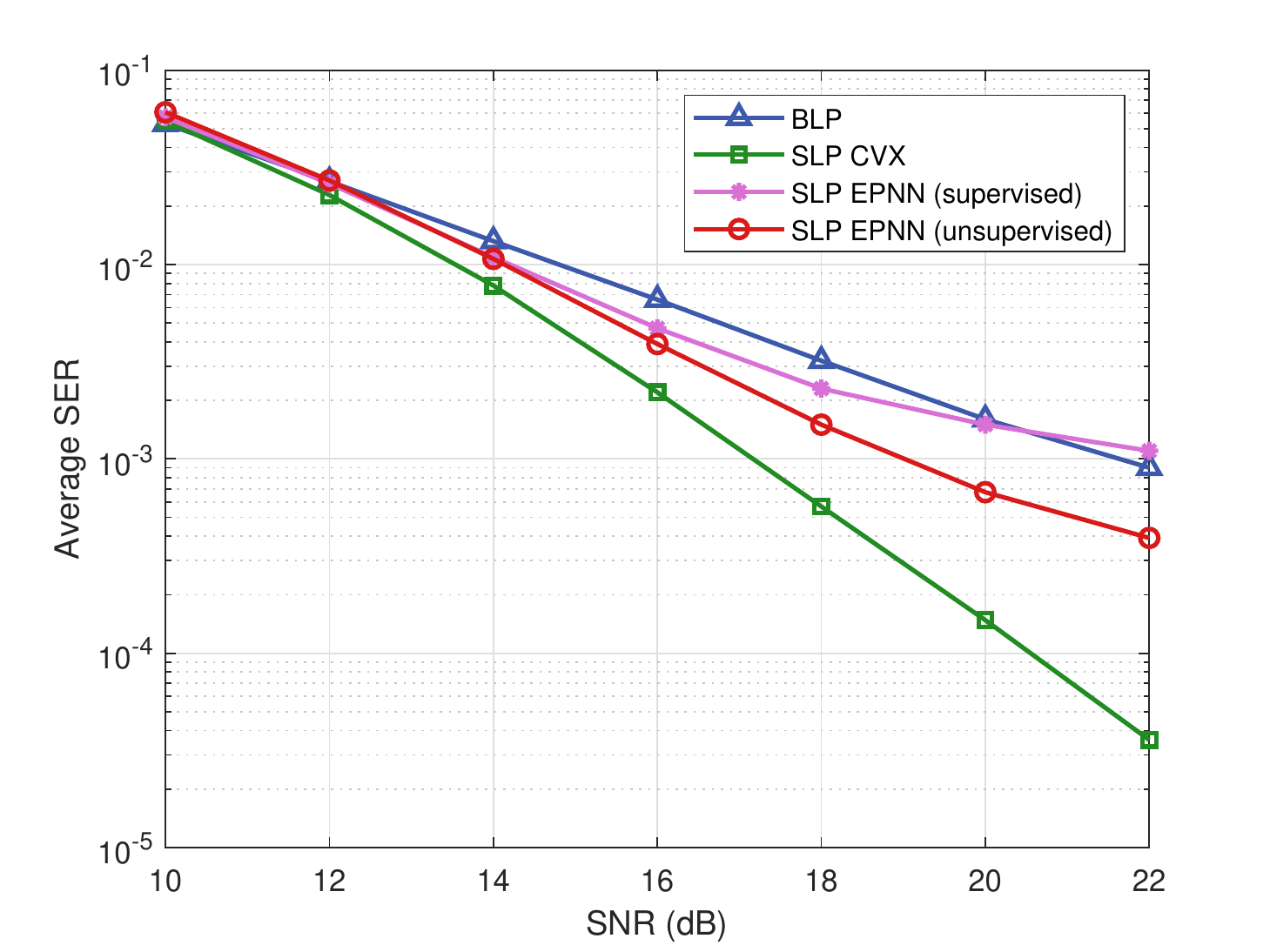}}
\vspace{0.3 cm}
\caption{Average SER versus SNR.}
\vspace{0.0 cm}
\end{figure}
We generate 4,000,000 realizations of $\mathbf{H}$ for training usage and 50,000 channel realizations for testing.
In the training stage, we train the network with 60 epoches by employing Adam optimizer.
In order to accelerate the training process and ensure a good performance, the initial learning rate is set as $\eta=0.001$ and the learning rate is decreased by the decay factor $\alpha=0.1$ every 20 epoches.
The mini-batch size is 1024 and the regularization factor $\lambda$ is fixed as 0.2 through the whole off-line training stage.
We utilize the proposed unsupervised scheme and the supervised scheme (MSE) to train the network.
For comparison purpose, we adopt CVX solver to design SLP as the non-deep-learning traditional approach, denoted by ``SLP CVX''.
All SLP schemes only design $N_\mathrm{par}$ precoding vectors for each channel realization.
The full precoding matrix will be obtained from the partial vectors by rotation operation.
Besides, the conventional block-level precoding algorithm in \cite{BLP} is adopted for comparison, which is denoted as ``BLP''.

We first illustrate the average SER versus transmit SNR for different precoding schemes in Fig. 3, where the $N_\mathrm{t} = 4, K = 3$ and $N_\mathrm{t} = 5, K = 4$ scenarios are shown in Figs. 3(a) and 3(b), respectively.
We see that both the ``SLP CVX'' and the proposed ``EPNN (unsuerpversied)'' approaches have better SER performance than the ``BLP'' scheme owing to the interference exploitation strategy of symbol-level precoding.
Furthermore, with the increase of SNR and the number of users, this advantage becomes more remarkable.
In addition, compared with the traditional ``SLP CVX'' approach, certain performance loss (less than 2dB) is observed in the proposed ``EPNN unsupervised'' approach, which is acceptable in practical implementations especially considering the significant computational complexity reduction.
Besides, the proposed ``EPNN (unsupervised)'' approach has better performance than ``EPNN (supervised)'' scheme.
This is because that the MSE metric, which is to minimize the Euclidean distance between $\widehat{\mathbf{X}}$ and the local optimal solutions $\mathbf{X}$ obtained by CVX, is irrelevant to the original optimization objective.
Specifically, the MSE based training scheme cannot always ensure that the MUI is constructive, i.e., pushing the signal as farther away from the decision boundaries as possible, and obtaining larger $d_{k,l}$.
Therefore, the MSE based supervised training scheme causes large performance loss.
In contrast, our proposed unsupervised scheme trains the network to
directly push the symbol away from the decision boundaries and thus obtains better SER performance.
Moreover, our proposed EPNN is not affected by the channel fading model, e.g., Rician fading model.

Finally, to provide a more intuitive and direct comparison of online design complexity, we compare the average running time of different SLP designs in Table II, where the traditional ``SLP CVX'' algorithm is run on CPU in Python environment for a fair comparison and the proposed EPNN is tested on both CPU and GPU.
It is worth noting that the training of EPNN is off-line.
The number of transmit antennas is set as $N_\mathrm{t}=5$.
It is obvious that the proposed EPNN dramatically reduces the running time on both CPU and GPU platforms.
Particularly, when $K = 5$, compared to traditional CVX method, our proposed EPNN takes only about 1.42\% and 0.15\% of the execution time on CPU and GPU, respectively.
Moreover, it can be observed that the running time of CVX method grows exponentially with the increasing number of users, while the proposed EPNN approach maintains almost the same running time, which reveals the effectiveness of the proposed DL-based SLP design for large-scale systems.
%\textcolor{blue}{It is worth noting that the one-shot off-line training complexity of EPNN is negligible in terms of average complexity, since it can be numerously applied for SLP design of flat fading time-varying channels.}
\begin{table}
\normalsize
\caption{Running time for different SLP schemes $(\mathrm{ms})$}
\begin{center}
\begin{tabular}{| c | c | c | c |}
\hline
& SLP CVX & \makecell{SLP EPNN \\(CPU)} & \makecell{SLP EPNN  \\(GPU)}\\
\hline 2 users&$2.968$  & $0.7368$ & $0.09354$ \\
\hline 3 users&$4.059$  & $0.8578$ & $0.1035$ \\
\hline 4 users&$9.095$  & $0.8681$ & $0.1134$ \\
\hline 5 users&$83.78$  & $1.186$  & $0.1256$ \\
\hline
\end{tabular}
\end{center}
%\vspace{-0.30 cm}
\end{table}

%\vspace{0.2cm}
\section{Conclusion}
\label{sc:Conclusions}
In this correspondence, we investigated the symbol-level precoding design to maximize the minimum quality-of-service (QoS) of all users with given power budget in the multi-user multi-input single-output (MU-MISO) system.
To avoid the huge computational complexity of symbol-level precoding design, we proposed a deep learning based approach that trains the efficient precoding neural network (EPNN) with an unsupervised-like method.
Simulation results verified the advantages of the proposed EPNN in dramatically reducing the computational complexity with satisfactory performance.

\end{document}